\documentclass[11pt,a4paper]{article}
\usepackage{jheppub}
\usepackage{amstext}
\usepackage{slashbox}

\title{Diquark contributions to Top quark charge asymmetry at the Tevatron and LHC}
\author{Kaoru Hagiwara,}
\author{Junya Nakamura}

\affiliation{KEK Theory Center and Sokendai, Tsukuba, Ibaraki 305-0801, Japan}

\emailAdd{kaoru.hagiwara@kek.jp}
\emailAdd{junnaka@post.kek.jp}

\abstract{We study contributions of a scalar diquark particle in a color representation of anti-triplet and sextet to the top quark pair production at the Tevatron and the Large hadron collider (LHC). The model can give Forward-Backward (FB) asymmetry at the Tevatron while can avoid the same sign top quark production at the LHC by assuming the top-number conserving diquark couplings. We study compatibility between the large positive FB asymmetry observed at the Tevatron and non-observation of the charge asymmetry at the LHC, by including contributions from the single and pair production of diquarks. We find that the whole parameter space of the models can soon be explored at the LHC by measuring the total $t\bar{t}$ production cross section and the inclusive charge asymmetry with smaller uncertainties. In addition, we compare the statistical significance of the charge asymmetry measured at the LHC with that of the optimal observable of the sub-process FB asymmetry, and find that they are comparable even when we ignore the uncertainty in the parton distribution functions.}


\begin{document}
\maketitle
\flushbottom

\section{Introduction}

The charge asymmetry in the $t\bar{t}$ production is the difference between the top and anti-top quark distributions. In the standard model (SM), small charge asymmetry is expected in the light quark pair annihilation processes, $q\bar{q}\to t\bar{t}$ at the next-to-leading order (NLO) of QCD perturbation theory \cite{chargeasym1}, which predicts small Forward-Backward (FB) asymmetry in $p\bar{p}$ collisions (Tevatron) and tiny charge asymmetry in $pp$ collisions (LHC).

At the Tevatron $t\bar{t}$ pair is produced mainly in the quark pair annihilation process, $q\bar{q}\to t\bar{t}$, where the quarks (anti-quarks) are mainly moving along the proton (anti-proton) momentum direction. The FB asymmetry can be defined as

\begin{align}
A_{\mathrm{FB}}=\frac{N(\Delta y>0)-N(\Delta y<0)}{N(\Delta y>0)+N(\Delta y<0)},\label{eq:afb}
\end{align}
where N is the number of events and $\Delta y=y_t-y_{\bar{t}}$ is the difference in rapidity of top and anti-top quarks along the proton momentum direction in the laboratory frame. 

Recent results from the CDF and the D0 collaboration at the Tevatron report positive asymmetries \cite{AfbCDF,AfbD0}
\begin{subequations}
\begin{align}
A^{\mathrm{CDF}}_{\mathrm{FB}}=15.8\pm 7.4\ \%  \label{eq:afbcdf}\\
A^{\mathrm{D0}}_{\mathrm{FB}}=19.6\pm 6.5\ \%   \label{eq:afbd0}
\end{align}
\end{subequations}
at the sub-process level after correcting for backgrounds and detector effects, while the SM prediction at the NLO in QCD is $5.0\pm 0.1\ \%$ \cite{AfbTeva1,AfbTeva2,AfbTeva3}. Furthermore, the CDF reported even larger asymmetry of $47.5\pm11.4$\% at $M_{t\bar{t}}>450$GeV \cite{AfbCDF} while the D0 reported the asymmetry of $15.2\pm4.0$\% in the rapidity of leptons \cite{AfbD0}, both exceeding the SM predictions by more than 3 standard deviations.

At the LHC, the sub-process FB asymmetry cannot be observed from $\Delta y=y_t-y_{\bar{t}}$ distribution because it is a $pp$ collider. Instead, the following charge asymmetry is sensitive to the $q\bar{q}\to t\bar{t}$ sub-process FB asymmetry at the LHC
\begin{align}
A_{\mathrm{C}}=\frac{N(\Delta |y|>0)-N(\Delta |y|<0)}{N(\Delta |y|>0)+N(\Delta |y|<0)},\label{eq:Ac}
\end{align}
where $\Delta |y|=|y_t|-|y_{\bar{t}}|$ is the difference in the absolute values of rapidities of top and anti-top quarks. The latest results from the ATLAS and the CMS collaboration at the LHC are \cite{AcATLAS,AcCMS}
\begin{subequations}
\begin{align}
A^{\mathrm{ATLAS}}_{\mathrm{C}}=-1.8\pm 2.8\ \mathrm{(stat.)}\pm 2.3\ \mathrm{(syst.)}\ \% , \label{eq:AcATLAS}\\
A^{\mathrm{CMS}}_{\mathrm{C}}=-1.3\pm 2.8\ \mathrm{(stat.)}^{\ +2.9}_{\ -3.1}\ \mathrm{(syst.)}\ \% ,\label{eq:AcCMS}
\end{align}
\end{subequations}
which are within the uncertainty consistent with the SM prediction, $A_C=1.15\pm0.06\ \%$ \cite{chargeasym2}. The asymmetry is small because the gluon fusion sub-process, $gg\to t\bar{t}$, gives the dominant contribution at the LHC. Nevertheless we expect the LHC data to reveal the asymmetry with the help of its high energy and its expected high luminosity.\\

In this paper, we study contributions of a scalar diquark particle in a color representation of anti-triplet and sextet to the total $t\bar{t}$ cross section and inclusive asymmetries at the Tevatron and the LHC. We evaluate the parameter space of the models which are consistent with the large positive FB asymmetry at the Tevatron, non-observation of the charge asymmetry at the LHC and non-deviation of the total $t\bar{t}$ production cross section from the SM prediction both at the Tevatron and at the LHC. 

The contribution of diquarks to the FB asymmetry at the Tevatron is discussed in refs.\cite{diquark1*,diquark2*,diquark3*,diquark4,diquark5*,diquark6*,diquark7,diquark8*,diquark9,diquark10*,
diquark11*,diquark12*,diquark16*,diquark17}
 and their contribution to the charge asymmetry at the LHC is discussed in refs.\cite{diquark13*,diquark14}. The $\phi$ single and pair production expected in the models is discussed in refs.\cite{diquark1*,diquark2*,diquark3*,diquark5*,diquark6*,diquark8*,diquark10*,diquark11*,
diquark12*,diquark13*,diquark15*,diquark16*}. We explore the parameter space of the models by including contributions from the single and pair production of diquarks to the charge asymmetry $A_C$ at the LHC.

This paper organized as follows. In the next section, we introduce the diquark models and discuss their phenomenological consequences at the Tevatron and the LHC. In Section 3, we show our numerical results of the diquark models. In Section 4, we introduce the optimal observable of the sub-process FB asymmetry. In Section 5, we compare the statistical significance of the charge asymmetry $A_C$ eq.(\ref{eq:Ac}) measured at the LHC with that of the optimal observable of the sub-process FB asymmetry, and examine the efficiency of $A_C$. The last section gives the summary.

\section{Diquark model}
We consider a model which consists of a new scalar boson $\phi$ in a SU(3) color representation of anti-triplet or sextet with a diquark quantum number:
\begin{align}
{\cal L} = \lambda \sqrt{2}C_{ij}^a t_R^i\cdot d_R^j \phi^{a*}+h.c.,\label{diquarklag}
\end{align}
where $C_{ij}^a$	is Clebsch-Gordon coefficients and described clearly in the appendix of ref.\cite{diquarkmodel}.
The representations under $SU(3)_C\times SU(2)_L\times U(1)_Y$ for anti-triplet and sextet diquarks are $(\bar{3},1,1/3)$ and $(6,1,1/3)$, respectively. We may denote the anti-triplet diquark as $\phi_{\bar{3}}=(\phi^1,\phi^2,\phi^3)^T$ and the sextet diquark as $\phi_{6}=(\phi^4,\phi^5,\phi^6,\phi^7,\phi^8,\phi^9)^T$. If only the interactions of eq.(\ref{diquarklag}) are present, the diquarks carry the quantum number of the top and down quarks and their chiralities, hence each flavor number as well as the chirality is conserved.

The SM gauge invariance allows top-light diquark couplings between the doublets $(t_L,b_L)$ and $(u_L,d_L)$, and between the singlets $t_R$ and $d_R$, or $u_R$. The diquarks of $t_R$ and $u_R$ have the same form of the Lagrangian eq.(\ref{diquarklag}) when $d_R^j$ is replaced by $u_R^j$ and $\phi^a$ have the hypercharge 4/3. The $SU(2)_L$ representation of the $(t_L,b_L)-(u_L,d_L)$ diquarks is either singlet or triplet, $\phi^{'}$ or $\phi^"=(\phi^"_{4/3}, \phi^"_{1/3}, \phi^"_{-2/3})^T$, respectively, whose Lagrangians are
\begin{subequations}\label{diquarklag4}
\begin{align}
{\cal L}_{\mathrm{SU(2)_L\ singlet}} &= \lambda^{'} \sqrt{2}C_{ij}^a \frac{1}{\sqrt{2}}(t_L^i\cdot d_L^j - b_L^i\cdot u_L^j)  \phi^{'a*}+h.c.,\label{diquarklag2}\\
{\cal L}_{\mathrm{SU(2)_L\ triplet}} &= \lambda^{"} \sqrt{2}C_{ij}^a \left\{-t_L^i\cdot u_L^j \phi^{"a*}_{4/3} + b_L^i\cdot d_L^j \phi^{"a*}_{-2/3} + \frac{1}{\sqrt{2}}(t_L^i\cdot d_L^j + b_L^i\cdot u_L^j)  \phi^{"a*}_{1/3}\right\}+h.c..\label{diquarklag3}
\end{align}
\end{subequations}
The electromagnetic (EM) charge 1/3 diquarks contribute to $d\bar{d}\to t\bar{t}$, whereas the EM charge 4/3 diquarks contribute to $u\bar{u}\to t\bar{t}$. We find that all the diquarks give positive FB asymmetry, especially at high $M_{t\bar{t}}$, because of the chirality conservation of each Lagrangian. The left-chirality diquarks of eq.(\ref{diquarklag4}) also contribute to the single top quark production, $u\bar{d}\to t\bar{b}$, and hence are strongly constrained \cite{singletopCMS,singletopATLAS}.

Rather than studying all of them exhaustively, we choose to study the $t_R$-$d_R$ diquarks of eq.(\ref{diquarklag}), since the anti-triplet diquark $\phi_{\bar{3}}$ has the least exotic quantum number among all the diquarks, which can be the super-partner of the right-handed down quark if the supersymmetry breaking scale is far above the electroweak scale \cite{rparity}.

The diquark interactions of eq.(\ref{diquarklag}) can generate the sub-process FB asymmetry at the leading order in the process of 
\begin{align}
d_i(p_1,\lambda_1)+\bar{d}_j(p_2,\lambda_2)\to t_k(p_3,\lambda_3)+\bar{t}_l(p_4,\lambda_4)\label{eq:process1}
\end{align}
through the {\it u}-channel exchange of $\phi$, as illustrated in Figure 1, where $p_i$ and $\lambda_i$ are momenta and helicities, respectively, and {\it i, j, k} and {\it l} are the color indices.

\begin{figure}[t]
\centering
\includegraphics[scale=0.4]{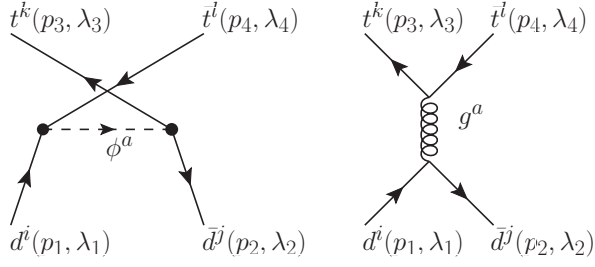}
\caption{Feynman diagrams of the process $d\bar{d}\to t\bar{t}$ through the {\it u}-channel exchange of $\phi^a_{\bar{3},6}$, in addition to the QCD amplitude with {\it s}-channel exchange of gluon ($g^a$). Hence {\it i, j, k, l} and {\it a} are the color indices, $p_n$ and $\lambda_{n}$ are momenta and helicities, respectively.}
\end{figure}

The helicity amplitudes $M^{kl\lambda_3\lambda_4}_{ij\lambda_1\lambda_2}$ are summarized below,
\begin{subequations}\label{helall}
\begin{align}
M_{ij+-}^{kl++}&=\frac{|\sqrt{2}\lambda|^2}{m_{\phi}^2-\hat{u}} \sum_a C_{il}^a C_{jk}^{a*} \frac{\sqrt{\hat{s}}}{2} m_t \sin{\theta}+\frac{4g^2}{\hat{s}} \sum_{a=1}^8  T_{ji}^{a} T_{kl}^a \frac{\sqrt{\hat{s}}}{2} m_t \sin{\theta}\label{hel1}\\
M_{ij+-}^{kl+-}&=\frac{|\sqrt{2}\lambda|^2}{m_{\phi}^2-\hat{u}} \sum_a C_{il}^a C_{jk}^{a*} \left(\frac{\sqrt{\hat{s}}}{2}\right)^2 (1+\beta) (1+\cos{\theta})+\frac{4g^2}{\hat{s}} \sum_{a=1}^8  T_{ji}^{a} T_{kl}^a \left(\frac{\sqrt{\hat{s}}}{2}\right)^2 (1+ \cos{\theta})\label{hel2}\\
M_{ij+-}^{kl-+}&=-\frac{|\sqrt{2}\lambda|^2}{m_{\phi}^2-\hat{u}} \sum_a C_{il}^a C_{jk}^{a*}  \left(\frac{\sqrt{\hat{s}}}{2}\right)^2 (1-\beta) (1-\cos{\theta})-\frac{4g^2}{\hat{s}} \sum_{a=1}^8  T_{ji}^{a} T_{kl}^a \left(\frac{\sqrt{\hat{s}}}{2}\right)^2 (1- \cos{\theta})\label{hel3}\\
M_{ij+-}^{kl--}&=-\frac{|\sqrt{2}\lambda|^2}{m_{\phi}^2-\hat{u}} \sum_a C_{il}^a C_{jk}^{a*}  \frac{\sqrt{\hat{s}}}{2} m_t \sin{\theta}-\frac{4g^2}{\hat{s}} \sum_{a=1}^8  T_{ji}^{a} T_{kl}^a \frac{\sqrt{\hat{s}}}{2} m_t \sin{\theta},\label{hel4}
\end{align}
\end{subequations}
where the first terms correspond to the amplitudes of the {\it u}-channel exchange of $\phi^a_{\bar{3}}$ or $\phi^a_{6}$ and the second terms give the QCD one-gluon exchange amplitudes. Initial down quarks are assumed massless, and hence the diquarks do not contribute to the $M_{ij-+}^{kl\lambda_3\lambda_4}$ amplitudes, which are obtained from the QCD part of the amplitudes eq.(\ref{helall}) by parity transformation ($\lambda_i \to -\lambda_i$, $\cos{\theta}\to -\cos{\theta}$). Here $\hat{s}=(p_1+p_2)^2$ is the invariant mass squared of the $t\bar{t}$ system, $\beta=(1-4m_t^2/\hat{s})^{1/2}$ is the velocity of the top quark, $\theta$ is the polar angle between the initial down quark and the final top quark momenta in the $t\bar{t}$ rest frame.

The color-space propagators of the $\bar{3}$ and $6$ diquarks as well as that of QCD gluons are, respectively, 
\begin{subequations}
\begin{align}
\sum_{a=1}^3 C_{il}^a C_{jk}^{a*}=\frac{1}{2}(\delta_{ji}\delta_{kl}-\delta_{jl}\delta_{ki}),\label{color1}\\
\sum_{a=4}^9 C_{il}^a C_{jk}^{a*}=\frac{1}{2}(\delta_{ji}\delta_{kl}+\delta_{jl}\delta_{ki}),\label{color2}\\
\sum_{a=1}^8  T_{ji}^{a} T_{kl}^a=\frac{1}{2}(\delta_{jl}\delta_{ki}-\frac{1}{3}\delta_{ji}\delta_{kl}),\label{color3}
\end{align}
\end{subequations}
and the color factors which appears in the color summed squared amplitudes are
\begin{subequations}
\begin{align}
\sum_{i,j,k,l=1}^3\sum_{a=1}^3\sum_{b=1}^8 (C_{il}^a C_{jk}^{a*})^* T^b_{kl}T^b_{ji}=- 2,\label{colorsum1}\\
\sum_{i,j,k,l=1}^3\sum_{a=4}^9\sum_{b=1}^8 (C_{il}^a C_{jk}^{a*})^* T^b_{kl}T^b_{ji}=+ 2,\label{colorsum2}\\
\sum_{i,j,k,l=1}^3\sum_{a,b=1}^3 (C_{il}^a C_{jk}^{a*})^* C_{il}^b C_{jk}^{b*}=3,\label{colorsum3}\\
\sum_{i,j,k,l=1}^3\sum_{a,b=4}^9 (C_{il}^a C_{jk}^{a*})^* C_{il}^b C_{jk}^{b*}=6,\label{colorsum4}\\
\sum_{i,j,k,l=1}^3\sum_{a,b=1}^8 (T^a_{kl}T^a_{ji})^* T^b_{kl}T^b_{ji}=2.\label{colorsum5}
\end{align}
\end{subequations}
At this stage, we can tell that the leading part of the QCD amplitudes interfere constructively (destructively) with the sextet (anti-triplet) diquark exchange amplitudes, and we expect positive FB asymmetry for the sextet contribution at all energies. Even for the anti-triplet, we find positive FB asymmetries for strong couplings $\lambda$ when the diquark exchange amplitude dominates over the QCD amplitude in eq.(\ref{hel2}), which happens e.g. at $\beta=0.46, 0.33, 0.14$ for $\lambda=2.6, 2.8, 3.0$ with $m_{\phi}=500$GeV, respectively. The positive FB asymmetry is a consequence of the chirality conservation of the effective Lagrangian eq.(\ref{diquarklag}), and hence it is common for all the diquark models that respect the SM gauge invariance as in eq.(\ref{diquarklag4}).

At the LHC, the positive FB asymmetry at the sub-process level gives rise to the positive charge asymmetry $A_C$ eq.(\ref{eq:Ac}). In addition, single and pair production of diquarks can be significant at the LHC,
\begin{subequations}
\begin{align}
dg\to \phi\bar{t} \to t\bar{t}d,\label{eq:process2}\\
\bar{d}g\to t\bar{\phi} \to t\bar{t}\bar{d},\label{eq:process1.5}\\
gg \to \phi\bar{\phi} \to t\bar{t}d\bar{d},\label{eq:process3}\\
q\bar{q} \to \phi\bar{\phi} \to t\bar{t}d\bar{d}.\label{eq:process4}
\end{align}
\end{subequations}
If the diquark $\phi$ is heavier than the top quark, $m_{\phi} > m_t$, the single production processes give $t\bar{t}+$jet events and the pair production processes give $t\bar{t}+2$jets events. These additional processes can also contribute to the inclusive FB and charge asymmetry as well as to the total $t\bar{t}$ production rate. The relevant diagrams for the single and pair production of diquarks via $dg\to\phi\bar{t}$ and $d\bar{d}\to \phi\bar{\phi}$ are shown in Figure 2 and Figure 3, respectively.

\begin{figure}[t]
\centering
\includegraphics[scale=0.4]{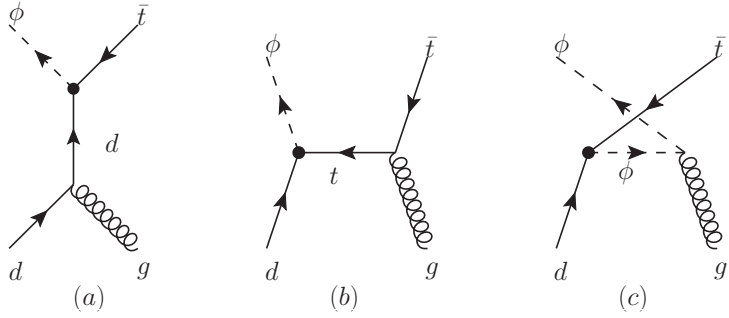}
\caption{Feynman diagrams of the process $dg\to \phi\bar{t}$.}
\end{figure}

\begin{figure}[t]
\centering
\includegraphics[scale=0.4]{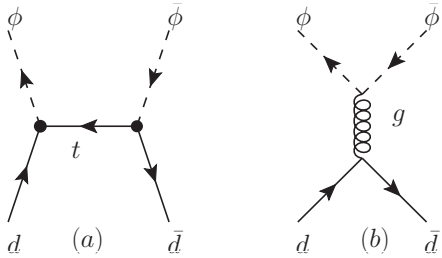}
\caption{Feynman diagrams of the process $d\bar{d} \to \phi\bar{\phi}$.}
\end{figure}



The single $\phi$ production process of eq.(\ref{eq:process2}) is dominated by the {\it t}-channel top quark exchange amplitude in Figure 2(b) when the diquark is heavier than the top quark, and hence the diquark is produced mainly along the initial down quark momentum direction. This leads to $y_t > y_{\bar{t}}$ at the Tevatron. At the LHC, initial down quarks in protons are more boosted on average than gluons, and hence the diquark is mainly produced at large $|y_{\phi}|$, which leads to $|y_{t}|>|y_{\bar{t}}|$.

Similarly in the single $\bar{\phi}$ production process of eq.(\ref{eq:process1.5}), the anti-diquark $\bar{\phi}$ is produced mainly along the initial anti-down quark momentum direction due to the top quark exchange amplitude. This also leads to $y_t > y_{\bar{t}}$ at the Tevatron. At the LHC, however, 
since anti-down quarks have only small energy fractions of incoming protons, the charge asymmetry from the single $\bar{\phi}$ production is negligibly small.

In the $\phi$ pair production process of eq.(\ref{eq:process4}), the {\it t}-channel top quark exchange amplitude in Figure 3(a) gives a sub-process FB asymmetry $y_{\phi} > y_{\bar{\phi}}$, which leads to $y_t > y_{\bar{t}}$ at the Tevatron and $|y_{t}|>|y_{\bar{t}}|$ at the LHC. The qualitative effect of the charge asymmetry at the LHC is, however, negligibly small because of the smallness of the $\bar{d}$ energy fraction in the proton. The $\phi$ pair production have contributions mainly to the total $t\bar{t}$ production cross section, mainly via the $gg\to \phi\bar{\phi}$ sub-process of eq.(\ref{eq:process3}) which gives no asymmetry.

From the above discussion, single and pair production of diquarks also contribute positively to the FB asymmetry at the Tevatron and to the charge asymmetry at the LHC, as well as to the total $t\bar{t}$ production rate.

\section{Numerical results of the diquark models}




\subsection{Total $t\bar{t}$ cross section and asymmetries at Tevatron and LHC}

In this section, we examine the parameter space of the diquark models and look for allowed regions which are consistent with the measurements both at the Tevatron and the LHC.

At the Tevatron, while producing a large positive FB asymmetry, it is important to keep the deviation of the total cross section of the $t\bar{t}$ production $\sigma(p\bar{p}\to t\bar{t}X)$ from the SM prediction small because it has been measured to agree with the SM prediction \cite{xsecCDF1,xsecCDF2,xsecCDF3,xsecD01,xsecD02}. Thus we require the following two conditions to be satisfied,
\begin{subequations}\label{tevaconditionall}
\begin{align}
A_{FB} > 5\%,\label{tevacondition1}\\
\frac{|\sigma_{SM+NP}-\sigma_{SM}|}{\sigma_{SM}}(p\bar{p}\to t\bar{t}X) < 0.2 \label{tevacondition2},
\end{align}
\end{subequations}
in the leading QCD order, and consider that the parameter region is excluded if one of the above conditions is not satisfied. Since the SM predicts $A_{FB}$ of about $5\%$ \cite{AfbTeva1,AfbTeva2,AfbTeva3} at the NLO of perturbative QCD, our requirement of $A_{FB}>5\%$ at the leading order may be large enough to be consistent with the large positive FB asymmetry observed at the Tevatron. The condition eq.(\ref{tevacondition2}) for the total cross section is rather loose, since the observed $t\bar{t}$ production cross sections \cite{xsecCDF1,xsecCDF2,xsecCDF3,xsecD01,xsecD02} tend to be somewhat larger than the NLO+NLL QCD prediction of about $7.14$ pb for $m_t=172.5$ GeV \cite{xsecTevaTheo}.

At the LHC, on the other hand, since the uncertainties on the $t\bar{t}$ total cross section and the charge asymmetry $A_C$ are still large, we use the observed upper bounds from the experimental data. As for the total cross section, we require \cite{xsecATLAS}
\begin{align}
\sigma(pp\to t\bar{t}X) < 204\ \mathrm{pb}\ \mathrm{at\ 95\%\ confidence\ level\ (C.L.)},\label{xsec95}
\end{align}
and for the charge asymmetry, we require one of the following constraints \cite{AcATLAS}
\begin{subequations}
\begin{align}
A_C<2.8\ \%\ \mathrm{at\ 90\%\ C.L.},\label{Ac90}\\
A_C<4.1\ \%\ \mathrm{at\ 95\%\ C.L.}.\label{Ac95}
\end{align}
\end{subequations}



We use CTEQ6L1 \cite{CTEQ} parton distribution function (PDF), and take the SM parameters as the top quark mass $m_{t}=172.5\ \mathrm{GeV}$ and the QCD coupling $\alpha_S(m_Z)=0.130$ which is used in CTEQ6L1, throughout this paper. All the results are obtained by using {\sc MadGraph} \cite{Madgraph, Madgraph2, Madgraph3} at the matrix element level. We do not take into account the effects from parton showering, hadronizations and detector conditions.

The result is shown in Figure 4. The horizontal axis corresponds to the diquark mass $m_{\phi}$ and the vertical axis to the coupling constant $\lambda$. The anti-triplet diquark model is evaluated in the top panels, and the sextet model is in the down panels. The two panels on the left hand side are obtained without contributions from the single and pair production of diquarks, and the two figures on the right hand side are obtained with contributions from the single and pair production of diquarks. Two dashed lines reflect the conditions at the Tevatron eq.(\ref{tevaconditionall}), and the shaded region satisfies both of these two conditions. The space on the right hand side (RHS) of the solid line satisfies both eqs.(\ref{xsec95}) and (\ref{Ac95}), and the space on the RHS of the dotted line satisfies both eq.(\ref{xsec95}) and the stronger constraint eq.(\ref{Ac90}).

\begin{figure}[t]
\centering
\includegraphics[scale=0.55]{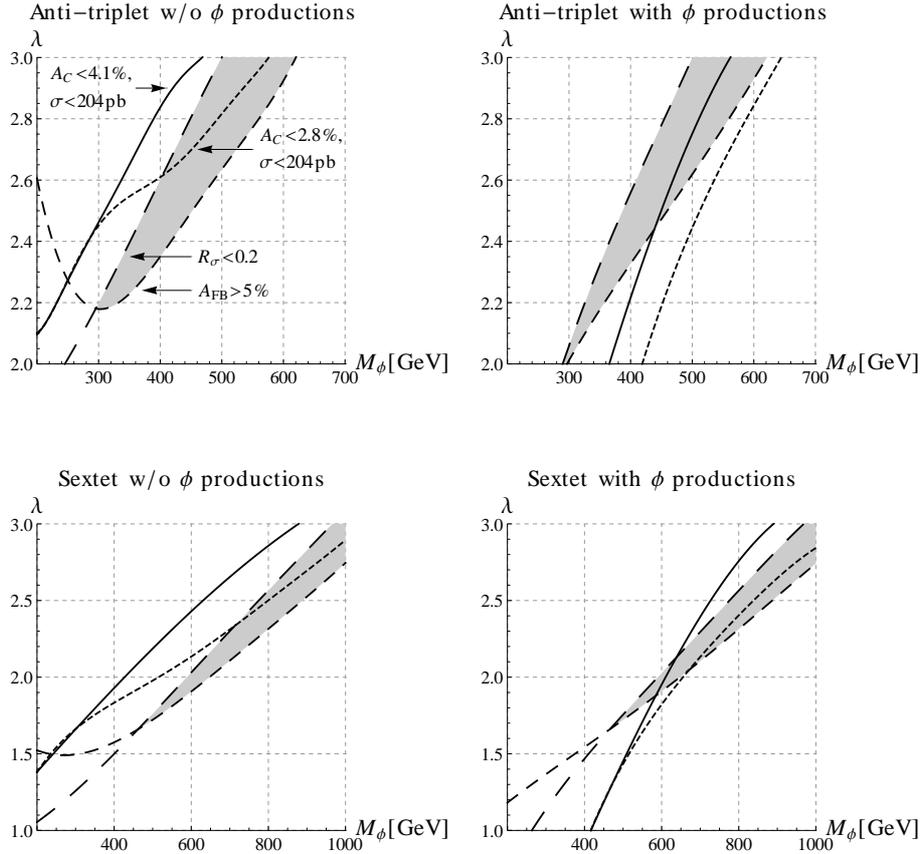}
\caption{The horizontal axis corresponds to the diquark mass $m_{\phi}$ and the vertical axis to the coupling constant $\lambda$. The anti-triplet diquark model is evaluated in the top panels, and the sextet model is in the down panels. The two panels on the left hand side are obtained without contributions from the single and pair production of diquarks, and the two panels on the right hand side are obtained with contributions from the single and pair production of diquarks. Two dashed lines reflect the conditions at the Tevatron eq.(\ref{tevaconditionall}), and the shaded region satisfies both of these two conditions. The space on the right hand side (RHS) of the solid line satisfies both eqs.(\ref{xsec95}) and (\ref{Ac95}), and the space on the RHS of the dotted line satisfies both eq.(\ref{xsec95}) and the stronger constraint eq.(\ref{Ac90}).}
\end{figure}

Comparing the left and the right panels in Figure 4, we find that the contribution from the single and pair production of diquarks is quite significant. The allowed region of the anti-triplet model is significantly reduced when $\phi$ production contributions are taken into account in the right panel. It is also worth noting that $\phi_{\bar{3}}$ production contributes to the positive FB asymmetry even at the Tevatron when $m_{\phi}<400$ GeV. To a lesser extent, $\phi_6$ production contributions reduce the allowed region of the sextet model as shown by the bottom two panels in Figure 4. We should therefore consider contributions from the single and pair production of diquarks, as well as from the {\it u}-channel diquark exchange process, when we discuss the FB and charge asymmetry in the diquark models.

The anti-triplet diquark model can be excluded if we adopt the constraint on $A_C$ at $90\%$ C.L. from the LHC eq.(\ref{Ac90}), whereas the sextet diquark model still has small compatible parameter space in the RHS of the dotted line in the shaded region of the bottom-right panel of Figure 4. If we loosen the constraint on $A_C$ to $95\%$ C.L. limit of eq.(\ref{Ac95}), both models still have allowed parameter spaces. Nevertheless, the allowed parameter regions are rather narrow, and can soon be explored by measuring the total $t\bar{t}$ production cross section and the inclusive charge asymmetry with smaller uncertainties at the LHC.

\subsection{Acceptances of the $t\bar{t}$ events induced by diquark productions}

In Section 3.1, we find that effects of the single and pair production rate of diquarks can be large at the LHC. When we discuss these contributions to $A_C$ and $\sigma(pp\to t\bar{t}X)$ in Figure 4, we assume that $100\%$ of the diquark production contribute to the inclusive $t\bar{t}$ events. However, in actual measurements of $A_C$ and $\sigma(pp\to t\bar{t}X)$, experimentalists apply event selection cuts on final states in order to increase the ratio of the number of the signal $t\bar{t}$ events over that of background events. Since the single and pair productions of the diquark lead to $t\bar{t}+1$ jet and $t\bar{t}+2$ jets events, respectively, at the sub-process level, the probability of those events to pass the selection cuts can differ from that of the SM $t\bar{t}$ events.

In this section, we study the acceptance of the $t\bar{t}$ events induced by the single and pair productions of the diquark by using Monte Carlo generated event samples. For this study, {\sc MadGraph/MadEvent} \cite{Madgraph, Madgraph2, Madgraph3} is used to obtain the parton level distributions for $t\bar{t}+n$ jets events, which are interfaced to {\sc Pythia6} \cite{Pythia} for the parton showering with the shower $k_{T}$ matching scheme \cite{showerkT}. We generate $t\bar{t}$ events where one of the W bosons decays into an electron or muon and the corresponding neutrino and the other W boson decays into a pair of jets. After jet clustering, we require jets to satisfy,
\begin{subequations}\label{jetcut}
\begin{align}
p_T^j&>30 \mathrm{\ GeV},\\
|\eta^j|&<2.5,\\
\mathrm{Number\ of\ jets}&\ge4.
\end{align}
\end{subequations}
We further impose the following selection cuts,
\begin{subequations}\label{electronch}
\begin{align}
p_{T}^e&>30 \mathrm{\ GeV},\\
|\eta^e|&<2.5,\\
E_{T}^{\mathrm{miss}}&>35 \mathrm{\ GeV},\\
m_{T}(l,\nu)&>25 \mathrm{\ GeV},\\
\frac{E^{\mathrm{Total}}_{\mathrm{Jet}}(\Delta R < 0.4)}{p_T^e}&<0.125,\label{electronch1}
\end{align}
\end{subequations}
for the electron channel, and 
\begin{subequations}\label{muonch}
\begin{align}
p_{T}^{\mu}&>20 \mathrm{\ GeV},\\
|\eta^{\mu}|&<2.1,\\
E_{T}^{\mathrm{miss}}&>20 \mathrm{\ GeV},\\
E_{T}^{\mathrm{miss}}+m_{T}(l,\nu)&>60 \mathrm{\ GeV},\\
\frac{E^{\mathrm{Total}}_{\mathrm{Jet}}(\Delta R < 0.4)}{p_T^{\mu}}&<0.125,\label{muonch1}
\end{align}
\end{subequations}
for the muon channel, where $m_{T}(l,\nu)$ is the transverse mass $\sqrt{2p_T^lE_{T}^{\mathrm{miss}}(1-\cos(\phi^l-\phi^{\mathrm{miss}}))}$ and $E^{\mathrm{Total}}_{\mathrm{Jet}}(\Delta R < 0.4)$ is the sum of jet energies in a cone with $\Delta R = \sqrt{\Delta \phi_{lj}^2+\Delta \eta_{lj}^2} = 0.4$ around the lepton track. These event selection cuts are obtained from refs.\cite{AcCMS,AcATLAS}, although we do not simulate detector effects in this paper. We simply define the acceptance as a ratio of the number of events that pass the above selection cuts eqs.(\ref{jetcut}, \ref{electronch}, \ref{muonch}) over the number of generated events. 
The acceptance for the SM $t\bar{t}$ events are found to be $0.45$ in our analysis.

\begin{table}[t]
\centering
\begin{tabular}{|c|c|c|c|c|c|}
\hline
\backslashbox{Process}{$m_{\phi}$[GeV]} & 200 & 400 & 600 & 800 & 1000\\
\hline
$\phi\bar{t}$, $t\bar{\phi}$ & 1.05 & 1.11 & 1.15 & 1.15 & 1.14\\
\hline
$\phi\bar{\phi}$ &  1.07  & 1.17    & 1.18 & 1.18 & 1.18 \\
\hline
\end{tabular}
\caption{Acceptances of the single and pair productions of the diquark normalized to that of the SM $t\bar{t}$ process for different masses of the diquark.}
\end{table}

We show in Table 1 the acceptances of the single and pair production processes normalized to that of the SM $t\bar{t}$ process for different masses of the diquark at $\sqrt{s}=7$ TeV. As the mass of the diquark grows, we expect that not only the down-quark jet but also the top quark from the diquark decays are boosted, resulting in hard jets and a hard lepton in the final state which make it easy for the event to pass the selection cuts. However, we find in Table 1 that the acceptance does not simply increase as the mass grows. The reason is as follows. For heavy diquark with mass $\gtrsim 600$ GeV, the top quark from its decay is highly boosted and when this top quark decays leptonically, a lepton and a bottom quark tend to be collinear. These events tend to be rejected by the present selection cuts for the lepton isolation, eqs.(\ref{electronch1}, \ref{muonch1}). We have confirmed that the acceptances increase almost linearly with the diquark mass when the lepton isolation cuts, eqs.(\ref{electronch1}, \ref{muonch1}), are removed. All the results in Table 1 are obtained for the color anti-triplet diquark. We find similar results for the color sextet diquark.

In order to take into account higher acceptances for the diquark production events in the analysis of Figure 4, we simply multiply the diquark production cross sections by the acceptance ratios in Table 1. The cross sections become larger and therefore the solid and dotted lines shift a bit toward the right in the two panels on the right hand side in Figure 4, resulting in slightly reduced allowed regions. However, because the acceptance ratios are only $5$ to $20 \%$ larger than the unity and also because the allowed regions in Figure 4 corresponds to the regions where the diquark production cross sections are small, we find that the results presented in Section 3.1 are not affected significantly by taking account of the difference in the acceptance of the diquark production events and the SM $t\bar{t}$ events.


\section{Projection method}

\subsection{General formula}

In this section, we would like to study the relationship between the sub-process FB asymmetry and the charge asymmetry $A_C$, eq.(\ref{eq:Ac}), which is measured at the LHC.

The number of top and anti-top pair production events in pp collisions can be described as
\begin{subequations}
\begin{align}
N(\tau,Y,y)=&{\cal L}\Bigl[D_{gg}(\tau,Y)\hat{\sigma}_{gg}(\tau,y)+D_{q\bar{q}}(\tau,Y)\hat{\sigma}_{q\bar{q}}(\tau,y)+D_{\bar{q}q}(\tau,Y)\hat{\sigma}_{\bar{q}q}(\tau,y)\Bigr]\label{eq:N_1}\\ 
=&{\cal L}\Bigl[D_{gg}(\tau,Y)\hat{\sigma}_{gg}(\tau,y)+D_S^{q\bar{q}}(\tau,Y)\hat{\sigma}_{S}^{q\bar{q}}(\tau,y) + D_A^{q\bar{q}}(\tau,Y)\hat{\sigma}_{A}^{q\bar{q}}(\tau,y)\Bigr],\label{eq:N}
\end{align}
\end{subequations}
where ${\cal L}$ is an integrated luminosity, $\hat{\sigma}_{ab}(\tau,y)$ are the $ab\to t\bar{t}$ sub-process cross sections, $\tau=x_1x_2$ gives the product of the momentum fractions of the colliding partons, $y$ is the rapidity of the top quark in the sub-process rest frame, and $Y$ is the rapidity of the $t\bar{t}$ system in the laboratory frame. In eq.(\ref{eq:N_1}), $D_{ab}(\tau,Y)$ are the products of PDFs,
\begin{subequations}
\begin{eqnarray}
D_{gg}(\tau,Y)=D_{g/p}(x_a)\times D_{g/p}(x_b)=D_{g/p}(\sqrt{\tau} e^{+Y})\times D_{g/p}(\sqrt{\tau} e^{-Y}),\label{eq:Dgg}\\
D_{q\bar{q}}(\tau,Y)=D_{q/p}(x_a)\times D_{\bar{q}/p}(x_b)=D_{q/p}(\sqrt{\tau} e^{+Y})\times D_{\bar{q}/p}(\sqrt{\tau} e^{-Y}),\label{eq:Dqqbar}\\
D_{\bar{q}q}(\tau,Y)=D_{\bar{q}/p}(x_a)\times D_{q/p}(x_b)=D_{\bar{q}/p}(\sqrt{\tau} e^{+Y})\times D_{q/p}(\sqrt{\tau} e^{-Y}).\label{eq:Dqbarq}
\end{eqnarray}
\end{subequations}
In eq.(\ref{eq:N}), $\hat{\sigma}_{S}^{q\bar{q}}$ and $\hat{\sigma}_{A}^{q\bar{q}}$ are, respectively, the symmetric and anti-symmetric part of the sub-process cross section about $y$, or equivalently symmetric and anti-symmetric under the exchange of colliding $q$ and $\bar{q}$ momentum direction,
\begin{subequations}
\begin{align}
\sigma_{q\bar{q}}(\tau,y)=\sigma_{\bar{q}q}(\tau,-y)=\sigma_{S}^{q\bar{q}}(\tau,y)+\sigma_{A}^{q\bar{q}}(\tau,y),\label{eq:xsecqqbar}\\
\sigma_{q\bar{q}}(\tau,-y)=\sigma_{\bar{q}q}(\tau,y)=\sigma_{S}^{q\bar{q}}(\tau,y)-\sigma_{A}^{q\bar{q}}(\tau,y).\label{eq:xsecqbarq}
\end{align}
\end{subequations}
Similarly, $D_S^{q\bar{q}}$ and $D_A^{q\bar{q}}$ in eq.(\ref{eq:N}) denote the symmetric and anti-symmetric combinations of $D_{q\bar{q}}(\tau,Y)$ in eq.(\ref{eq:Dqqbar}) and $D_{\bar{q}q}(\tau,Y)$ in eq.(\ref{eq:Dqbarq}),
\begin{subequations}
\begin{align}
D_S^{q\bar{q}}(\tau,Y)=D_{q\bar{q}}(\tau,Y)+D_{\bar{q}q}(\tau,Y),\label{eq:Ds}\\
D_A^{q\bar{q}}(\tau,Y)=D_{q\bar{q}}(\tau,Y)-D_{\bar{q}q}(\tau,Y).\label{eq:Da}
\end{align}
\end{subequations}
In Figure 5, we show the $Y$ distribution of the product of the $d$ and $\bar{d}$ quark PDF's; the black solid curve gives $D_{d\bar{d}}(\tau,Y)$ of eq.(\ref{eq:Dqqbar}), the black dashed curve gives $D_{\bar{d}d}(\tau,Y)$ of eq.(\ref{eq:Dqbarq}), whereas the blue and red curves are their symmetric $D_S^{d\bar{d}}(\tau,Y)$ of eq.(\ref{eq:Ds}) and anti-symmetric $D_A^{d\bar{d}}(\tau,Y)$ of eq.(\ref{eq:Da}), respectively.

\begin{figure}[t]
\centering
\includegraphics[scale=0.6]{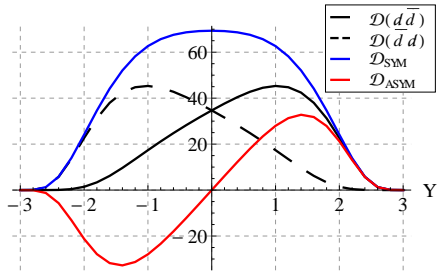}
\caption{The $Y$ distribution of the product of the $d$ and $\bar{d}$ quark PDF's at 7 TeV LHC; the black solid curve gives $D_{d\bar{d}}(\tau,Y)$ of eq.(\ref{eq:Dqqbar}), the black dashed curve gives $D_{\bar{d}d}(\tau,Y)$ of eq.(\ref{eq:Dqbarq}), whereas the blue and red curves are their symmetric $D_S^{d\bar{d}}(\tau,Y)$ of eq.(\ref{eq:Ds}) and anti-symmetric $D_A^{d\bar{d}}(\tau,Y)$ of eq.(\ref{eq:Da}), respectively. The product of the momentum fractions of the colliding partons, $\tau$, is set to $\tau=4m_t^2/s$, and the factorization scale is set to $m_t$. CTEQ6L1 PDF is used \cite{CTEQ}.}
\end{figure}

In eq.(\ref{eq:N}), multiplying both sides by $D_A^{q\bar{q}}(\tau,Y)$ and integrating about $Y$ in its whole region, the first and second terms on the right hand side vanish because these terms become anti-symmetric about $Y$, and only the third term survives,
\begin{align}
\int_{all}dY\int_{\tau_j}^{\tau_{j+1}}d\tau D_A^{q\bar{q}}(\tau,Y)N(\tau,Y,y)
={\cal L}\int_{all}dY\int_{\tau_j}^{\tau_{j+1}}d\tau\bigl(D_A^{q\bar{q}}(\tau,Y)\bigr)^2\hat{\sigma}_A^{q\bar{q}}(\tau,y).\label{eq:proj}
\end{align}
In other words, the asymmetric part of the sub-process cross section, $\hat{\sigma}_A^{q\bar{q}}(\tau,y)$, can be projected out even in pp collisions if we know the $q$ and $\bar{q}$ PDF accurately. In the following sections, we examine the possibility of the optimal measurement of the asymmetric cross section of the $q\bar{q}\to t\bar{t}$ sub-process in the absence of the PDF uncertainty.

Making an approximation that $\hat{\sigma}_A^{q\bar{q}}(\tau_j)$ is constant between $\tau_j$ and $\tau_{j+1}$, we obtain the asymmetric part of a sub-process cross section $\hat{\sigma}^{q\bar{q}}(\tau_j)$ as follows
\begin{eqnarray}
\hat{\sigma}_A^{q\bar{q}}(\tau_j)=\frac{\int_{all}dY\int_{\tau_j}^{\tau_{j+1}}d\tau \int_{all}dy \bigl[\theta(y)-\theta(-y)\bigr] D_A^{q\bar{q}}(\tau,Y)N(\tau,Y,y)}{{\cal L}\int_{all}dY\int_{\tau_j}^{\tau_{j+1}}d\tau\bigl(D_A^{q\bar{q}}(\tau,Y)\bigr)^2}.\label{eq:xsecA2}
\end{eqnarray}
Since there is no statistical correlation between $N(\tau,Y,y)\theta(y)$ and $N(\tau,Y,y)\theta(-y)$, the statistical uncertainty of $\hat{\sigma}_A^{q\bar{q}}(\tau_j)$ is estimated as,
\begin{eqnarray}
\Delta \hat{\sigma}_A^{q\bar{q}}(\tau_j)=\frac{\sqrt{\int_{all}dY\int_{\tau_j}^{\tau_{j+1}}d\tau \int_{all}dy \bigl[\theta(y)+\theta(-y)\bigr] \bigl(D_A^{q\bar{q}}(\tau,Y)\bigr)^2N(\tau,Y,y)}}{{\cal L}\int_{all}dY\int_{\tau_j}^{\tau_{j+1}}d\tau\bigl(D_A^{q\bar{q}}(\tau,Y)\bigr)^2}.\label{eq:errorxsecA}
\end{eqnarray}
If we can ignore the PDF uncertainty in $D_A^{q\bar{q}}(\tau,Y)$, the projected cross section eq.(\ref{eq:xsecA2}) should have the smallest error since the weight $D_A^{q\bar{q}}(\tau,Y)$ minimizes its statistical uncertainty \cite{optimal1,optimal2,optimal3,optimal4}.

\begin{figure}[t]
\centering
\includegraphics[scale=0.6]{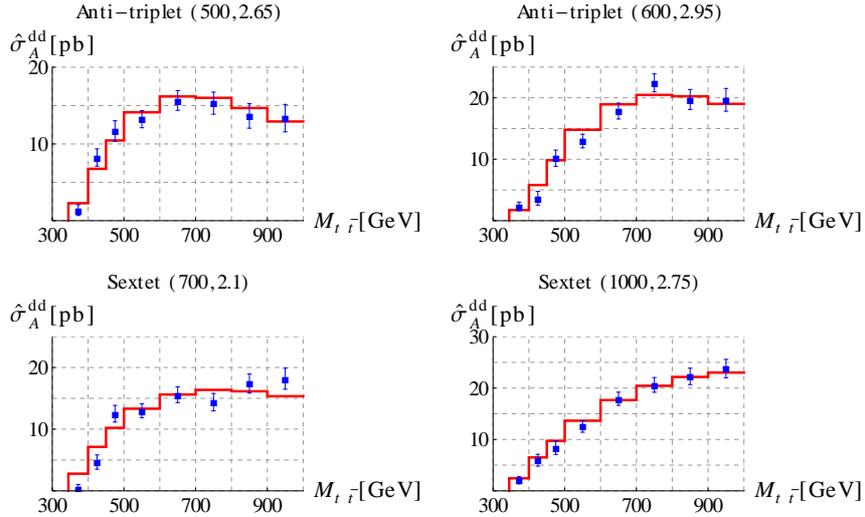}
\caption{The results of extracting $\hat{\sigma}_{A}^{d\bar{d}}$ from MC generated $t\bar{t}$ event samples as a function of the $t\bar{t}$ invariant mass $M_{t\bar{t}}$, with the statistical uncertainties. The theoretical calculations are shown as red solid curves. The color anti-triplet (top panels) and sextet diquark (bottom panels) models are assumed. Parameter sets for the diquark mass $m_{\phi}$ and the coupling constant $\lambda$ are shown above each panel.}
\end{figure}

\subsection{Extracting $\hat{\sigma}_{A}^{q\bar{q}}(\tau)$ from MC generated event samples}  
In this section, we demonstrate the extraction of $\hat{\sigma}_{A}^{q\bar{q}}(\tau)$ from Monte Carlo (MC) generated $t\bar{t}$ event samples by using eq.(\ref{eq:xsecA2}). Event samples are generated by {\sc MadGraph/MadEvent} \cite{Madgraph, Madgraph2, Madgraph3} assuming the color anti-triplet and sextet diquark models. {\sc FeynRules} is used to make the model files for {\sc MadGraph} \cite{feynrules}. From the allowed parameter spaces in Figure 4, we choose two points, $(m_{\phi_{\bar{3}}},\lambda)=(500,2.65)$ and $(600,2.95)$ for the anti-triplet model, $(m_{\phi_6},\lambda)=(700,2.1)$ and $(1000,2.75)$ for the sextet model. Since the diquark model eq.(\ref{diquarklag}) gives the sub-process FB asymmetry at the leading order in the process of $d\bar{d}\to t\bar{t}$ eq.(\ref{eq:process1}) through the {\it u}-channel exchange of the diquark, only the $d\bar{d}$ contribution, $\hat{\sigma}_{A}^{d\bar{d}}(\tau)$, is non-zero at the leading QCD order, and this is what we try to extract from MC generated $t\bar{t}$ event samples by using eq.(\ref{eq:xsecA2}). The MC generated event samples correspond to an integrated luminosity of $20\mathrm{fb^{-1}}$ at the 7 TeV LHC and we take into account the overall QCD NLO K factor of $1.5$ and the branching ratio 0.29 of the semi-leptonic $t\bar{t}$ decay into $l\nu +$jets for $l=e$ or $\mu$.

The results of extracting $\hat{\sigma}_{A}^{d\bar{d}}(\tau)$ with the statistical uncertainties are shown in Figure 6. The theoretical calculations of $\hat{\sigma}_{A}^{d\bar{d}}(\tau)$ are shown as red solid curves. The horizontal axis is the $t\bar{t}$ invariant mass, $M_{t\bar{t}}=\sqrt{\tau s}$. The results are consistent with the theoretical calculations within the statistical uncertainties, and hence we confirm that the projection formula eq.(\ref{eq:xsecA2}) works.

\section{Evaluation of the charge asymmetry $A_C$}
The charge asymmetry $A_C$ of eq.(\ref{eq:Ac}) measured at the LHC uses the difference $\Delta |y|=|y_t|-|y_{\bar{t}}|$ in the magnitudes of the $t$ and $\bar{t}$ rapidities along the proton momentum direction in the laboratory frame. In terms of $Y$ and $y$, $|y_t|$ and $|y_{\bar{t}}|$ are written as 
\begin{subequations}
\begin{eqnarray}
|y_t|=|Y+y|=(Y+y)\theta(Y+y)-(Y+y)\theta(-(Y+y)),\label{eq:y1}\\
|y_{\bar{t}}|=|Y-y|=(Y-y)\theta(Y-y)-(Y-y)\theta(-(Y-y)).\label{eq:y2}
\end{eqnarray}
\end{subequations}
Therefore, the numerator of the charge asymmetry eq.(\ref{eq:Ac}) can be expressed as
\begin{align}
N(\Delta |y|>0)-N(\Delta |y|<0)=\int_{all} d\tau \int_{all} dY \int_{all} dy\ N(\tau,Y,y)\times \nonumber\\
\biggl\{\Bigl[\theta(y)-\theta(-y)\Bigr]\Bigl[\theta(Y-y)\theta(Y+y)-\theta(-Y+y)\theta(-Y-y)\Bigr]\nonumber \\
+\Bigl[\theta(Y)-\theta(-Y)\Bigr]\Bigl[\theta(-Y+y)\theta(Y+y)-\theta(Y-y)\theta(-Y-y)\Bigr]\biggr\}.\label{eq:N11}
\end{align}
We note here that, if we simply define
\begin{align}
\Delta |y|'=(y_t-y_{\bar{t}})\theta(Y)-(y_t-y_{\bar{t}})\theta(-Y),\label{eq:delyprime}
\end{align}
instead of $\Delta |y|=|y_t|-|y_{\bar{t}}|$, then the numerator of eq.(\ref{eq:Ac}) becomes, instead of eq.(\ref{eq:N11}),
\begin{align}
N(\Delta |y|'>0)-N(\Delta |y|'<0)
=\int_{all} d\tau \int_{all} dY \int_{all} dy\ N(\tau,Y,y)\times \nonumber\\
\Bigl[\theta(y)-\theta(-y)\Bigr]\Bigl[\theta(Y)-\theta(-Y)\Bigr],\label{eq:N23}
\end{align}
and we confirm that inclusive values of eq.(\ref{eq:N11}) and eq.(\ref{eq:N23}) are numerically the same, although y- and Y- distributions are different. From eq.(\ref{eq:N23}), we find that the numerator of the charge asymmetry $A_C$ eq.(\ref{eq:Ac}) can be obtained by the projection method of eq.(\ref{eq:xsecA2}) with the weight function of 
\begin{align}
w(\tau,Y)=\theta(Y)-\theta(-Y),\label{weight}
\end{align}
instead of the optimal weight function $D_A^{q\bar{q}}(\tau,Y)$.

In order to compare the efficiency of the charge asymmetry $A_C$ eq.(\ref{eq:Ac}) and that of the optimal observable eq.(\ref{eq:xsecA2}), we introduce a variable $N_A^{Opt.}$,
\begin{align}
N_A^{Opt.}(\tau_j)\equiv
\int_{\tau_j}^{\tau_{j+1}}d\tau \int_{all}dY \int_{all} dy\ \Bigl[\theta(y)-\theta(-y)\Bigr]D_A^{q\bar{q}}(\tau,Y)N(\tau,Y,y),\label{eq:Na11}
\end{align}
and define the optimal charge asymmetry $A_C^{Opt.}$ as
\begin{align}
A_C^{Opt.}(\tau_j)\equiv \frac{N_A^{Opt.}(\tau_j)}{N_{Total}(\tau_j)}.\label{eq:Ac31}
\end{align}

We evaluate the efficiency of these two observables by using MC generated $t\bar{t}$ event samples assuming the anti-triplet diquark model of eq.(\ref{diquarklag}) with $m_{\phi}=500$ GeV and $\lambda = 2.65$. The result is shown in Figure 7. The blue diamonds give the charge asymmetry $A_C$ of eq.(\ref{eq:Ac}) whereas the green squares give the optimal observable $A_C^{Opt.}$ defined by eq.(\ref{eq:Ac31}). The left panel shows the charge asymmetries as functions of the invariant mass of the $t\bar{t}$ system, $M_{t\bar{t}}$. The right panel shows the distributions of the significance, Asymmetry/Error, also as functions of $M_{t\bar{t}}$. Only statistical errors are considered. In both panels, each bin size of $M_{t\bar{t}}$ is $100$ GeV and the last bin includes all contribution of $M_{t\bar{t}}\ge1300$ GeV. The statistical error of $N_A^{Opt.}(\tau_j)$ is estimated as
\begin{align}
\Delta N_A^{Opt.}(\tau_j)=\sqrt{\int_{\tau_j}^{\tau_{j+1}}d\tau \int_{all}dY \int_{all} dy\ \Bigl[\theta(y)+\theta(-y)\Bigr]\Bigl(D_A^{q\bar{q}}(\tau,Y)\Bigr)^2N(\tau,Y,y)}.\label{eq:Naerror}
\end{align}

Although the significance of the optimal observable $A^{Opt.}_C$ is better than that of the charge asymmetry $A_C$ in all regions of $M_{t\bar{t}}$ as expected, they are almost comparable. We therefore conclude that the charge asymmetry $A_C$ is a very sensitive observable of the sub-process FB asymmetry at the LHC and that it is difficult to require it further once the PDF uncertainty are taken into account.

\begin{figure}[t]
\centering
\includegraphics[scale=0.7]{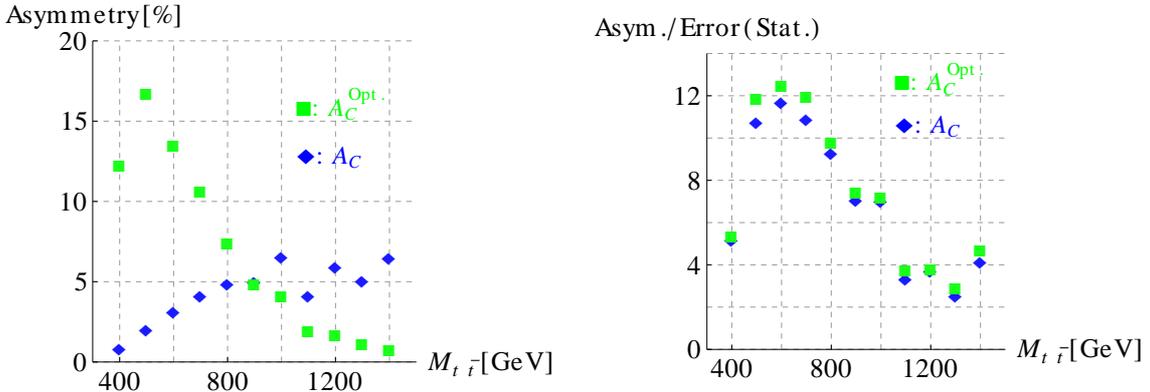}
\caption{The charge asymmetries (left) and their significances (right) as functions of the $t\bar{t}$ invariant mass, $M_{t\bar{t}}$. The blue diamonds give the charge asymmetry of eq.(\ref{eq:Ac}) whereas the green squares give the optimal observable defined by eq.(\ref{eq:Ac31}). Only statistical errors are considered. In both figures, each bin size of $M_{t\bar{t}}$ is $100$GeV and the last bin includes all contribution of $M_{t\bar{t}}\ge1300$ GeV.}
\end{figure}

\section{Conclusion}
In this paper, we have studied contributions of a scalar diquark particle in a color representation of anti-triplet and sextet to the top quark production at the Tevatron and the LHC. 

We have explored the parameter space of the diquark models which are consistent with the measurements both at the Tevatron and at the LHC by including contributions from the single and pair production of diquarks. We find that the sextet diquark model has a wider allowed parameter space than the anti-triplet diquark model does, and that the anti-triplet model can be excluded by the constraint on the charge asymmetry $A_C$ at $90\%$ C.L. from the LHC. At $95$\% C.L., both models still have allowed parameter spaces, which can soon be explored at the LHC.

In addition, we have introduced the optimal observable of the sub-process FB asymmetry and compared its statistical significance with that of the charge asymmetry $A_C$ measured at the LHC. We find that they are comparable and that $A_C$ is a very sensitive measure of the FB asymmetry in the $q\bar{q}\to t\bar{t}$ sub-processes.

\acknowledgments{We would like to thank the organizers and lecturers at the KIAS school on MadGraph for LHC physics simulation (Oct. 24-29, 2011, KIAS, Soul) and at the 2011 IPMU-YITP School and Workshop on Monte Carlo Tools for LHC (Sept. 5-10, 2011, Kyoto University, Kyoto) where we learned basics of hadron collider physics, which are applied in this study. J.N. thanks KIAS for financial support to attend the KIAS school. J.N. also thanks Yoshitaro Takaesu  and Kentarou Mawatari for valuable discussions and their help with {\sc MadGraph}. We are thankful that the simulation in this study was executed on the Central Computing System of KEK. The work is supported in part by Grant-in-Aid for Scientific Research (\#20340064 and \#23104006) from JSPS.}

\end{document}